\documentstyle[aps,prl,psfig,epsfig,twocolumn]{revtex}
\input{epsf}
\draft
\begin{document}
\title{Broken symmetries and directed collective energy transport}
\author{S. Flach$^\dag$, Y. Zolotaryuk$^\ddag$, A. E.
Miroshnichenko$^\dag$
and M. V. Fistul$^\dag$}
\address{
$^\dag$
Max-Planck-Institut f\"ur Physik komplexer Systeme, N\"othnitzer
Strasse 38, D-01187 Dresden, Germany \\
$^\ddag$ Section for Mathematical Physics, IMM, Technical
University of Denmark, \\DK-2800, Kgs. Lyngby, Denmark }

\date{\today}
\wideabs{
\maketitle
\begin{abstract}
We study the appearance of directed energy current in
{\it homogeneous} spatially extended systems
coupled to a heat bath in the presence of an external
ac field $E(t)$. The systems are described by nonlinear field equations.
By making use of a symmetry analysis we  predict the
right choice of $E(t)$ and obtain directed energy transport
for systems with a nonzero topological charge $Q$.
We demonstrate that the symmetry properties of motion of topological
solitons (kinks and antikinks) are equivalent to the ones
for the energy current.
Numerical simulations confirm the predictions
of the symmetry analysis and, moreover, show that the
directed energy current drastically increases as the dissipation parameter
$\alpha$ reduces.
Our results generalize recent rigorous theories of
currents generated by broken time-space symmetries to the case of
interacting many-particle systems.
\end{abstract}
}

The idea of rectifying energy transport with the help of fluctuations
has been discussed for several years in connection with
molecular motors and other nonequilibrium properties of
biological systems \cite{fjaajp97},
electrical currents in superlattices \cite{mix}, voltages in
Josephson junction coupled systems \cite{jj1}, to name a few.
The fluctuations have zero mean value, i. e. the dc component is absent.
The reduction of the underlying
problem, namely directed energy transport,
to a particle moving in a space-periodic but
asymmetric (ratchet) potential allowed to study the resulting
{\it directed current} in great details \cite{rbphjgk94} (for a recent review

see \cite{pr00}). A recently elaborated symmetry
approach to this problem  established a
clear relationship between directed currents and broken
space-time symmetries \cite{sfoyyz00,oysfyzaao01}. The essential step
was to separate the unavoidable correlations in the fluctuations
from the uncorrelated ones.
This is easily obtained by replacing the fluctuations as
a superposition of ac driving fields and uncorrelated white noise.
An important consequence is
that the symmetries may be broken either by violating the
reflection symmetry of the potential in space or by violating
the shift symmetry of the ac fields. Thus,  a particle may display
a directed motion also in the case of a space-symmetric potential.
Another interesting result is the persistence of directed currents
in the Hamiltonian limit of systems exposed to ac fields but decoupled
from the heat bath \cite{sfoyyz00,oysfyzaao01}.

A very important question is whether the symmetry approach can
be generalized to the case of interacting many-particle systems
(for related studies see \cite{pr00,scfflmf01,zzghbh01,msyz01}).
In this Letter we focus on nonlinear partial differential equations,
which can be considered as classical analogues of quantum models
of interacting particles \cite{rr87}.
We  show that such systems allow for a directed energy transport
when being driven by a proper combination of ac forces.
The success of such
a generalization will not only underline the general validity
of the symmetry analysis, but will also allow for a systematic
symmetry analysis of nonadiabatic nonlinear response functionals
for many-body theories.

We  study the properties of nonlinear Klein-Gordon equations
for a scalar field $\varphi$ which depends both on a spatial coordinate
$x$ and time $t$:
\begin{eqnarray}
&& \varphi_{,tt}-c_0^2\varphi_{,xx}+
\nonumber
\alpha\varphi_{,t}+\frac{d U}{d \varphi}=f(t,x)~,~~\\
\label{eq1}
&& f(t,x)=E(t)+\xi(t,x)~,
\end{eqnarray}
where $c_0$ is the limiting propagation speed of
small amplitude plane waves,
$\alpha$ is the dissipation parameter determining the
inverse relaxation time in the system,
and $f_{,z} \equiv \partial f / \partial z$.
The ac field $E(t)$ has zero mean and period $T$. The
Gaussian white noise
$\xi$ is characterized by the standard correlation function
$\langle \xi(t,x) \xi(t',x') \rangle = 2\frac{\alpha}{\beta}
\delta(x-x') \delta (t-t')$ where $\beta$ is the inverse temperature.
The potential $U(z)$ is assumed to have several identical minima for
$z=z_i$, $i=1,2,...$. We choose fixed boundary conditions:
$\varphi(x\rightarrow -\infty) = z_l$, $\varphi(x\rightarrow +\infty)
= z_m$. These boundary conditions determine
the {\it topological charge} $Q=m-l$ which will be of importance in the
following.
By making use of
the standard continuity equation $\rho_{,t} + j_{,x}=0$ , where
the energy density $\rho = (\varphi_{,t}^2+c_0^2\varphi_{,x}^2)/2+U(
\varphi)$, we obtain the following expression for the
energy current $J$:
\begin{equation}
J(t) =
\int_{-\infty}^{+\infty} j \; dx =
- c_0^2 \int_{-\infty}^{+\infty} \varphi_{,t} \varphi_{,x} dx~.
\label{flux}
\end{equation}

Next we turn to the symmetry analysis of (\ref{eq1},\ref{flux}).
It is carried out
in the absence of $\xi$ in
(\ref{eq1}), as it does not affect the symmetries.
We note that $J$ changes its sign under the transformation $x
\rightarrow -x$.
Eq. (\ref{eq1}) is invariant under the space inversion.
However the only topological charge invariant under
this transformation is $Q=0$. Consequently we
conclude that for zero topological charge, $Q=0$, the
time-averaged energy current $\langle J \rangle $
vanishes exactly.

In order to study the possibility of the appearance of a nonzero energy
current in the presence of a nonzero topological charge,
$Q\neq 0$, we note that the topological
charge $Q\neq 0$ is invariant under the combined symmetry operation
$x \rightarrow -x$ and $\varphi \rightarrow -\varphi + z_l + z_m$.
The same symmetry operation changes the sign of the energy current, and all
we
need is to ensure that the
equation (\ref{eq1}) remains invariant. This is now only possible if
the ac field possesses shift symmetry
and the potential $U(z)$ is symmetric around $z=(z_l+z_m)/2$.
Thus, we arrive at the symmetry operation
\begin{eqnarray}
x\rightarrow -x\;,\;\varphi \rightarrow -\varphi + z_l + z_m\;,\;
t \rightarrow t+\frac{T}{2}~,
\label{sym}
\end{eqnarray}
that leaves the equation (\ref{eq1}) invariant and changes the sign
of the energy current as the constraints
\begin{eqnarray} \label{cond}
&& E(t)= -E(t+T/2)~, \nonumber\\
&& U\left( z-\frac{z_l+z_m}{2}\right)=U\left( -z
+\frac{z_l+z_m}{2}\right)~,
\end{eqnarray}
hold.
In such a case we may conclude that the directed energy current vanishes
\cite{comment}.
Violating (\ref{cond}) we loose the symmetry (\ref{sym}) and
consequently may expect a nonzero energy current provided no further
hidden symmetries are overlooked.

Most interesting is that the conditions (\ref{cond})
can be violated by choosing
a function $E(t)$ which is not shift symmetric. A simple
choice
is
\begin{equation}
E(t)=E_1 \cos \omega t + E_2 \cos (2\omega t + \Delta)~,
\label{et}
\end{equation}
with $E_1 \neq 0$ and $E_2 \neq 0$.
In particular, such an ac field allows to generate a
nonzero energy current in the well
known
sine-Gordon equation with
\begin{equation}
U(z) = -\cos z~,
\label{sg}
\end{equation}
which we will consider in the following.
At the same time we stress here that any other choice of
the ac drive $E(t)$ which is not shift symmetric (e.g. a corresponding
sequence of pulses) will do the job, as well as any other
potential $U(z)$ which meets the above described conditions.

Of special interest is that in the
Hamiltonian limit as $\alpha \rightarrow 0$, time reversal
symmetry is recovered: the symmetry operation
$t \rightarrow -t$ leaves (\ref{eq1})
invariant, provided $E(t)$ is symmetric: $E(t) = E(-t)$.
As time reversal always changes the sign of the energy current
(\ref{flux}), zero average
energy current will be a consequence.
Thus, for the underdamped case $\alpha \ll 1$ we expect a
change of sign
of the energy current
upon varying $\Delta$, because for the particular values of $\Delta=0,\pi$
time reversal is approximately
restored and the directed energy current disappears.

As discussed above, a necessary condition for a directed energy current
is a nonzero topological charge $Q$. It implies
the presence of an excess of kinks over antikinks or vice versa.
The simplest case is $Q=1$  as a single kink is present on average
in the system. As a kink possesses a nonzero rest energy $E_k = 8 c_0$,
it is natural
to expect that the nonzero energy current is generated by a directed
motion of such a kink. Indeed, assuming that the kink position
$X$ and velocity $V$
are described by expressions
\begin{equation}
X(t) = \frac{1}{2\pi Q} \int_{-\infty}^{+\infty} x \varphi_{,x} dx\;,\;
V = \frac{1}{2\pi Q} \int_{-\infty}^{+\infty} x \varphi_{,xt} dx~.
\label{kinkpos}
\end{equation}
we observe that the symmetry operation (\ref{sym}) changes the sign of $V$.
Thus, we find that if the energy current vanishes by symmetry,
the same is true for the average velocity of a kink. This shows the
intimate connection between a directed energy current and
coherent kink excitations as
carriers of such an energy.
In \cite{msyz01} a directed kink motion was obtained numerically
for the first time
for the case (\ref{et}),(\ref{sg}) together with a study of its
complex dynamics.
Note that in \cite{fm96,sq01,avsgptavz97} a
directed kink transport has been obtained by breaking the
reflection symmetry of $U(z)$ for $E_2=0$.

If the temperature and thus the amplitude of $\xi$ is large enough,
additional kink-antikink pairs are excited in the course
of evolution. Since the space inversion ($x \rightarrow -x$)
does not change eq. (\ref{eq1}), transforms $Q \rightarrow -Q$
and kinks into antikinks with inverted
velocities, the average velocity of a kink-antikink pair is zero.
Consequently the generation of
kink-antikink pairs does not affect the net energy current of the
system predefined by the value of topological charge $Q$.

With these analytical results we turn to a numerical investigation.
We used a standard discretization scheme and integrated the following
coupled differential equations:
\begin{equation}
\ddot{\phi}_n - c_0^2\left( \phi_{n-1} - 2\phi_n
+ \phi_{n+1}\right) + \alpha \dot{\phi}_n
+ \frac{ d U}{ d \phi_n} = f(t)~.
\end{equation}
Here $n$ is an integer representing the discretization in space.
We choose the following parameters: $\beta = 100$,
$c_0^2 = 10$,
$E_1 = 0.2$, $E_2 = 0.2$, $\omega = 0.1$ and $Q=1$.
This particular value of $c_0$ ensures that  the effects of discreteness
on the kink motion may be neglected \cite{sfcrw93}.
%
\begin{figure}[htb]
\vspace{20pt}
\centerline{\psfig{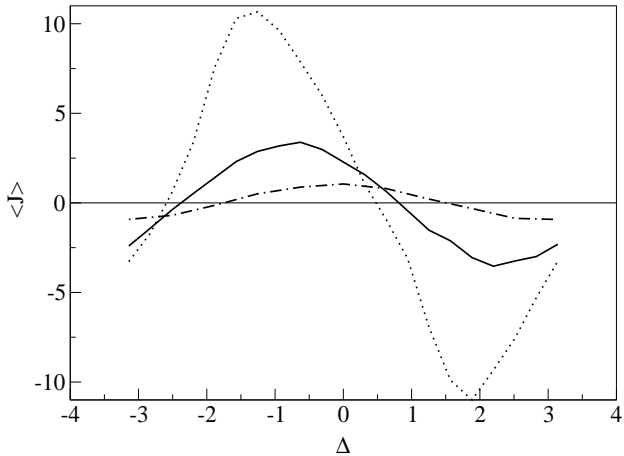}}
\vspace{2pt}
\caption{Dependence of the time-averaged
energy current $\langle J \rangle$
on the phase shift $\Delta$.
Solid line - $\alpha=0.05$, dotted line - $\alpha=0.01$,
dashed-dotted line - $\alpha=0.2$. Note that the latter case
$\alpha=0.2$ is scaled by a factor of 5, so the real values
of $\langle J \rangle$ are five times less than they appear in
the plot.}
\label{fig1}
\end{figure}
The time averaged energy current
$\langle J \rangle $ as
a function of the phase shift $\Delta$ for three different
values of dissipation parameter
$\alpha=0.01,0.05,0.2$ is shown in Fig. 1.
We obtain a strong dependence of the directed energy current
$\langle J \rangle $ on $\Delta$
with sign changes, as expected.
Furthermore the maximum values of the energy current increase
substantially as the dissipation parameter $\alpha$ reduces,
similar to the case of a single particle moving in a space-periodic
potential \cite{oysfyzaao01}. Especially we find an increase by a factor
of 50 when the damping $\alpha$ decreases from 0.2 to 0.01.
And finally
we clearly observe that as the parameter $\alpha$ reduces, the zeroes
of the curves tend to the positions determined by time reversal
symmetry, $\Delta~=~0,\pi$.
We computed in a similar way the time averages of $V$ as defined
in (\ref{kinkpos}), and obtained identical results upon variation
of $\Delta$ and $\alpha$. It shows that indeed kink motion is
responsible for the energy transfer.

In these simulations the amplitude of the white noise $\xi$ was
small enough to prevent
an excitation of a kink-antikink pair during the simulation time.
Nevertheless a very long simulation should also show up with
a spontaneous creation of such pairs. In order to see that this
event does not influence the symmetry analysis and especially
does not change the obtained values of the energy current (cf.
above considerations) we proceed  with a comparison.
In Fig. 2 we plot the dependence of the time-averaged energy current
$\langle J \rangle$ on the time of averaging for $\alpha=0.01 $
and $\Delta=\pi/2$, as a single kink is present
in the system. In the same figure we plot the result for the
case of additionally exciting a kink-antikink pair\cite{aemsvdaavts00}.
We indeed observe
that the value for the time-averaged energy current does not depend
on the number of additional kink-antikink pairs. As the simulation
time increases even further, soliton
collisions eventually lead to the dissipation of the pair,
with one single kink remaining in the system.
%
\begin{figure}[htb]
\vspace{20pt}
\centerline{\psfig{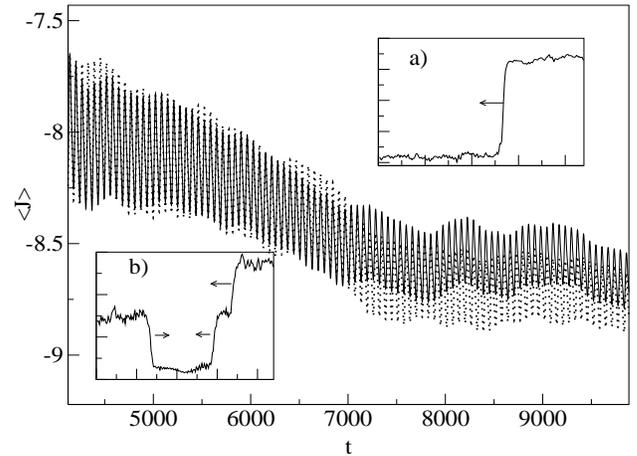}}
\vspace{2pt}
\caption{Dependence of the time-averaged
energy current $\langle J \rangle$ on the averaging time $t$
for $\Delta=\pi/2$ and $\alpha=0.01$ for the two cases: a single kink in the
system (solid line),  a single kink and an additional kink-antikink pair
in the system (dotted line). The insets (a) and (b) show snapshots of the
field $\varphi$ as a function of $x$ at some time during the simulation
for these two cases. Arrows indicate the direction of time-averaged motion
of the kink(s) and the antikink.}
\label{fig2}
\end{figure}
Let us discuss possible mechanisms of energy current rectification
in {\it spatially homogeneous} extended systems
through directed kink motion. First we notice that in many cases the
kink motion may be mapped to the classical mechanics of a macroscopic
particle
\cite{ms-msacs82}. However, a free particle that is subject to a spatially
homogeneous ac force does not display a directed motion at all.
One possible source of the observed
directed current can be relativistic effects which are inherent
to the considered nonlinear Klein-Gordon field equations.
They can be explicitly obtained in some cases using
projection
techniques to derive effective equations of motion for
the kink center \cite{ms-msacs82}.
In such a regime the kink motion is mapped onto the one of
a free relativistic particle in the presence of damping and
the ac field $E(t)$. It is straightforward to see that this
case leads to zero rectification for the limit of small damping,
in sharp contrast with our findings.
Thus we argue that the origin of the
observed strong rectification in the underdamped limit
is due to the nonadiabatic excitation of
{\it internal} kink modes and their interaction with the translational
kink motion \cite{msyz01,sq01,mvfegavu01}.

It is worthwhile to mention a number of other
ways to generate a nonzero energy current.
One case is to consider
strongly discrete systems where the appearance of a
nonnegligible Peierls-Nabarro potential can generate a net
kink motion
through the lattice.
Also all sorts of perturbations
in the form of spatial modulations may act similar to discretization
effects \cite{Carap,Gold}. Yet another way of breaking the symmetry has been
analyzed in \cite{sp95prl} where in addition to an ac force $E(t)=\cos
t$
the potential $U(z,t)=(z^2-1)^2 + z^2 \sin t $ was assumed to be
time-dependent.
The application
of our symmetry analysis yields that the symmetries (\ref{sym})
are violated and directed energy transport occurs.
It is also interesting to recall that a number of publications
have been dealing with directed kink motion in asymmetric double well
potentials \cite{avsgptavz97,gcfm01}. The asymmetry
was chosen in a way to
entropically favour one equilibrium over the other one (still
keeping the energy values of the equilibria identical). Even Gaussian
white noise leads then to the directed motion of a kink. However
additional ac fields have been shown to counterbalance the entropic
force.

While it is certain that our results may be applied to
many different physical situations where the sine-Gordon
equation is replaced by other Klein-Gordon equations
with topological excitations, we stress that
a particular system, namely an annular long Josephson junction with
a trapped fluxon, is perhaps the ideal case to verify our results.
This system is described precisely by (\ref{eq1})
\cite{mvfegavu01,GronbEn,FistUst}.
While in \cite{mvfegavu01} experimental results for $E_2=0$ have
been presented, all one needs is to add the second harmonics $E_2 \neq
0$
and vary the phase shift $\Delta$. The analogue of the ac field is here
an ac homogeneous microwave radiation which traverses the junction.
Under these conditions a trapped fluxon displays a directed motion,
and correspondingly a dc voltage drop across the junction appears.
This effect is the analogue of the energy current discussed here.
Notice here that the dc
voltage drop oscillates with the phase shift $\Delta$ and substantially
increases in the "strong" underdamped limit, $\alpha~\ll~\omega$.

We conclude this work with stating that the use of symmetries
allowed to obtain a relationship between a directed
energy current and directed kink motion. This may be of further
interest as it shows an intimate connection between symmetry
breakings and the relevance of coherent excitations
and their properties in
many body theories. Finally we note that nonlinear Klein-Gordon
systems which do not admit kink or antikink excitations
(zero
topological charge) do not allow for directed energy currents
induced by ac forces,
and that the above considerations can be used without
change for spatially discrete systems.

We thank M. Salerno and A. V. Ustinov for useful discussions.
This work was supported by the Deutsche Forschungsgemeinschaft
FL200/8-1 and the European Union through LOCNET HPRN-CT-1999-00163.
\bibliographystyle{unsrt}

\begin{thebibliography}{10}

\bibitem{fjaajp97}
F.~J\"ulicher, A.~Ajdari, and J.~Prost,
\newblock { Rev. Mod. Phys.} {\bf 69},  1269 (1997).

\bibitem{mix}
K. Seeger and W. Maurer,
\newblock{ Solid State Commun.}, {\bf 27}, 603 (1978);
I. Goychuk and P. H\"anggi,
\newblock { Europhys. Lett.}, {\bf 43}, 503 (1998);
K. N.~Alekseev {\it et al},
\newblock { Phys. Rev. Lett.}, {\bf 80}, 2669 (1998);
K. N.~Alekseev and F. V.~Kusmartsev,
{cond-mat/0012348}.

\bibitem{jj1}
I.~Zapata, J.~ \L uczka, F.~Sols, and P.~H\"anggi,
\newblock { Phys. Rev. Lett.} {\bf 80}, 829 (1998);
S. Weiss, D. Koelle, J. M\"uller, R. Gross, and K. Barthel,
\newblock{Europhys. Lett.} {\bf 51}, 499 (2000).

\bibitem{rbphjgk94}
R.~Bartussek, P.~H\"anggi, and J.~G. Kissner,
\newblock { Europhys. Lett.}, {\bf 28}, 459 (1994).

\bibitem{pr00}
P.~Reimann.
\newblock Brownian motors: noisy transport far from equilibrium.
\newblock { cond-mat/0010237}, Phys. Rep., in print.

\bibitem{sfoyyz00}
 S.~Flach, O.~Yevtushenko, and Y.~Zolotaryuk,
\newblock { Phys. Rev. Lett.} {\bf 84}, 2358 (2000).

\bibitem{oysfyzaao01}
O.~Yevtushenko, S.~Flach, Y.~Zolotaryuk, and A.~A. Ovchinnikov,
\newblock { Europhys. Lett.}, {\bf 54}, 141 (2001).

\bibitem{scfflmf01}
S.~Cilla, F.~Falo, and L.~M. Floria,
\newblock { Phys. Rev. E}, {\bf 63}, 031110 (2001).

\bibitem{zzghbh01}
Z.~Zheng, G.~Hu, and B.~Hu,
\newblock { Phys. Rev. Lett.}, {\bf 86}, 2273 (2001).

\bibitem{msyz01}
M.~Salerno and Y.~Zolotaryuk,
\newblock {preprint}.

\bibitem{rr87}
R.~Rajaraman,
\newblock {\em Solitons and Instantons}.
\newblock Elsevier, 1987.

\bibitem{comment} Notice here that even if the conditions (\ref{cond})
are satisfied a nonzero energy current may persist for a rather large
finite time (see for example B. A. Malomed and A. V. Ustinov, Phys. Rev. B
{\bf 64}, 020302 (2001). However, in this case the  energy current of
an opposite sign also exists, and a proper averaging over
different realizations leads to  zero total directed current.

\bibitem{fm96}
F. Marchesoni, \newblock{ Phys. Rev. Lett.} {\bf 77}, 2364 (1996).

\bibitem{sq01}
M.~Salerno and N.~R.~Quintero, {\it Soliton ratchets}, submitted to
Phys. Rev. Lett., (2001), cond-mat/0107011.

\bibitem{avsgptavz97}
 A.~V.~Savin, G.~P.~Tsironis, and A.~V. Zolotaryuk,
\newblock { Phys. Rev. E}, {\bf 56}, 2457 (1997).

\bibitem{sfcrw93}
S.~Flach and C.~R. Willis,
\newblock { Phys. Rev E}, {\bf 47}, 4447 (1993).

\bibitem{aemsvdaavts00} A kink-antikink pair was introduced in the system
through the exact three-soliton
solution of the unperturbed sine-Gordon equation, see
 A.~E. Miroshnichenko, S.~V. Dmitriev, A.~A. Vasiliev, and
T. Shigenari,
Nonlinearity {\bf 13}, 837 (2000).

\bibitem{ms-msacs82}
D. W.~McLaughlin and A. C.~Scott,
\newblock{Phys. Rev. A}, {\bf 18}, 1652 (1978);
M.~Salerno and A.~C. Scott,
\newblock  {Phys. Rev. B}, {\bf 26}, 2474 (1982).

\bibitem{mvfegavu01}
M.~V.~Fistul, E.~Goldobin,  and A.~V. Ustinov,
\newblock { Phys. Rev. B}, {\bf 64}, 092501 (2001).

\bibitem{Carap} G. Carapella and G. Costabile, Phys. Rev. Lett. {\bf 87},
077002 (2001).

\bibitem{Gold} E. Goldobin, A. Sterck, and D. Koelle, Phys. Rev. E {\bf 63},
031111 (2001).

\bibitem{sp95prl}
A. L. Sukstanskii and K. I. Primak,
\newblock{\em Phys. Rev. Lett.}, {\bf 75}, 3029 (1995).

\bibitem{gcfm01}
G.~Constantini and F.~Marchesoni,
\newblock { Phys. Rev. Lett.}, {\bf 87}, 114102 (2001).

\bibitem{GronbEn}
N. Gr{\o}nbech-Jensen, P. S. Lomdahl, and M. R. Samuelsen,
Phys. Rev. B {\bf 43}, 12799 (1991).

\bibitem{FistUst} M. V. Fistul and A. V. Ustinov,
Phys. Rev. B {\bf 63}, 024508 (2000).



\end{thebibliography}

\end{document}